\documentstyle[12pt]{article}
\begin{document}
\tolerance=5000
\def\pp{{\, \mid \hskip -1.5mm =}}
\def\cL{{\cal L}}
\def\be{\begin{equation}}
\def\ee{\end{equation}}
\def\bea{\begin{eqnarray}}
\def\eea{\end{eqnarray}}
\def\tr{{\rm tr}\, }
\def\nn{\nonumber \\}
\def\e{{\rm e}}
\def\D{{D \hskip -3mm /\,}}

\  \hfill
\begin{minipage}{3.5cm}
June 2001 \\
\end{minipage}

\vfill

\begin{center}
{\large\bf Conformal anomaly from dS/CFT correspondence}

\vfill

{\sc Shin'ichi NOJIRI}\footnote{nojiri@cc.nda.ac.jp},
and {\sc Sergei D. ODINTSOV}$^{\spadesuit}$\footnote{
odintsov@ifug5.ugto.mx, odintsov@mail.tomsknet.ru}\\

\vfill

{\sl Department of Applied Physics \\
National Defence Academy,
Hashirimizu Yokosuka 239-8686, JAPAN}

\vfill

{\sl $\spadesuit$
Tomsk State Pedagogical University, 634041 Tomsk, RUSSIA \\
and \\
Instituto de Fisica de la Universidad de Guanajuato,
Lomas del Bosque 103, Apdo. Postal E-143, 
37150 Leon,Gto., MEXICO 
}

\vfill

{\bf ABSTRACT}

\end{center}

In frames of dS/CFT correspondence suggested by Strominger
we calculate holographic conformal anomaly for dual euclidean CFT.
The holographic renormalization group method is used for this purpose.
It is explicitly demonstrated that two-dimensional and four-dimensional
conformal anomalies (or corresponding central charges) have the same form
as those obtained in AdS/CFT
duality.

\newpage

AdS/CFT duality (for a review, see \cite{AdS}) relates 
quantum gravity on AdS D-dimensional space with 
boundary CFT in one dimension less. It has been proven to be
very useful in modern string/quantum gravity/conformal theories.
It is suspected that AdS/CFT correspondence is manifestation
of some more deep fundamental principle somehow related with holography.

In the recent paper by Strominger \cite{strominger}(for related discussion,
see \cite{mottola}) it has been suggested 
dS/CFT correspondence in the similar sense as above AdS/CFT
correspondence. The fact that quantum gravity in de Sitter space 
may have some holographic dual has been already mentioned in several papers 
\cite{ds1,emil,ds2,ds3,ds4}. The following simple arguments may be suggested.
The reason why AdS/CFT can be expected is the isometry of 
$d+1$-dimensional anti-de Sitter space, which is $SO(d,2)$ 
symmetry. It is identical with the conformal symmetry of 
$d$-dimensional Minkowski space. We should note, however, the 
$d+1$-dimensional de Sitter space has the isometry of 
$SO(d+1,1)$ symmetry, which can be a conformal symmetry of 
$d$-dimensional Euclidean space. Then it might be natural to 
expect the correspondence between $d+1$-dimensional de Sitter 
space and $d$-dimensional euclidean conformal symmetry (dS/CFT 
correspondence\cite{strominger}). In fact, the metric of 
$D=d+1$-dimensional 
anti de Sitter space (AdS) is given by
\be
\label{AdSm}
ds_{\rm AdS}^2=dr^2 + \e^{2r}\left(-dt^2 + \sum_{i=1}^{d-1}
\left(dx^i\right)^2\right)\ .
\ee
In the above expression, the boundary of AdS lies at 
$r=\infty$.  
If one exchanges the radial coordinate $r$ and the time 
coordinate $t$, we obtain the metric of the de Sitter space (dS): 
\be
\label{dSm}
ds_{\rm dS}^2=-dt^2 + \e^{2t}\sum_{i=1}^d
\left(dx^i\right)^2\ .
\ee
Here $x^d=r$. Then there is a boundary at $t=\infty$, where the 
Euclidean conformal field theory (CFT) can live and one expects 
 dS/CFT correspondence.
The purpose of the present note is to get holographic conformal anomaly
within dS/CFT correspondence in the same style as it was done
in AdS/CFT correspondence (see refs. \cite{anom} for calculation of
holographic conformal anomaly in AdS/CFT duality). It is shown that
obtained central charge of dual CFT in two and four-dimensions is the same
as in the correspondent calculation in AdS/CFT. 

In the present discussion one can use the analogy with the
holographic renormalization group formulation
developed  in
\cite{BVV,FMS}. One can start from $D=d+1$ dimensional dS-like 
metric in the following form 
\be
\label{met}
ds^{2} = G_{MN}dX^{M}dX^{N} = 
-dt^{2} +\hat G_{\mu\nu}(x,t)dx^{\mu}dx^{\nu} .
\ee
where $X^{M}=(x^{\mu},t)$ with $\mu,\nu=1,2,\cdots ,d$. 
The gravitational action on a $(d+1)$ dimensional manifold $M_{d+1}$ with
the
boundary $\Sigma_{d}=\partial M_{d+1}$ is given by
\bea
\label{i}
S_{d+1}&=& \int_{M_{d+1}} d^{d+1}x\sqrt{-G}(R-\Lambda) + 
2\int_{\Sigma_{d}} d^{d}x \sqrt{\hat G}K \nn
&=& \int_{\Sigma_{d}}d^{d}x \int dr \sqrt{-G}\left( 
R-\Lambda - K_{\mu\nu}K^{\mu\nu} + K^{2} \right) \nn
&\equiv & \int d^{d}x dr \sqrt{-G} {\cal L}_{d+1}.
\eea
where $R$ and $K_{\mu\nu}$ are the scalar curvature and 
the extrinsic curvature on $\Sigma_{d}$ respectively.
We should note that we are considering the action in the 
Minkowski signature. 
Since we are considering the de Sitter background instead of 
the AdS background, the cosmological constant $\Lambda$ is 
positive and parametrized by the parameter $l$, which is the 
radius of the asymptotic dS$_{d+1}$
\be
\label{dSL}
\Lambda={d(d-1) \over l^2}\ .
\ee
$K_{\mu\nu}$ is given as
\bea
\label{ii}
K_{\mu\nu}={1\over 2}{\partial \hat G_{\mu\nu} 
\over \partial t},\quad
K=\hat G^{\mu\nu}K_{\mu\nu}
\eea
In the canonical formalism, ${\cal L}_{d+1}$ is rewritten
by using the canonical momenta $\Pi_{\mu\nu}$
and Hamiltonian density ${\cal H}$ as
\be
\label{iii}
{\cal L}_{d+1} = \Pi^{\mu\nu}{\partial 
\hat G_{\mu\nu} \over \partial t}
+{\cal H} \ ,\quad
{\cal H} \equiv {1 \over d-1}(\Pi ^{\mu}_{\mu})^2-\Pi_{\mu\nu}^{2}
+R-\Lambda \ . 
\ee
The equation of motion for $\Pi^{\mu\nu}$ leads to 
\bea
\label{iv}
\Pi^{\mu\nu}=K^{\mu\nu}-\hat G^{\mu\nu}K .
\eea
The Hamilton constraint ${\cal H}=0$ leads to the
Hamilton-Jacobi equation (flow equation) 
\bea
\label{HJ}
\{ S,S \} (x) &=& \sqrt{\hat G} {\cal L}_{d} (x) \\
\{ S,S \} (x) &\equiv & {1 \over \sqrt{\hat G}}
\left[-{1 \over d-1}\left(\hat G_{\mu\nu}{\delta S 
\over \delta \hat G_{\mu\nu}}
 \right)^{2}+\left( {\delta S \over \delta \hat G_{\mu\nu}}
 \right)^{2} \right] , \\
{\cal L}_{d}(x) &\equiv & R[\hat G] - \Lambda.
\eea
One can decompose the action $S$ into a local and non-local part as
discussed in ref.\cite{BVV} as follows
\bea
\label{v}
S[\hat G(x)] &=& S_{loc}[\hat G(x)]+\Gamma[\hat G(x)] ,
\eea
Here $S_{loc}[\hat G(x)]$ is tree level action and $\Gamma$ contains
the higher-derivative and non-local terms. 
In the following discussion, we take the systematic method 
of ref.\cite{FMS}, which is weight calculation. 
The $S_{loc}[\hat G]$ can be expressed as a sum of local terms
\be
\label{via}
S_{loc}[\hat G(x)] = \int d^{d}x \sqrt{\hat G} {\cal L}_{loc}(x) 
= \int d^{d}x \sqrt{\hat G} 
\sum _{w=0,2,4,\cdots} [{\cal L}_{loc}(x)]_{w} 
\ee
The weight $w$ is defined by following rules;
\[
\hat G_{\mu\nu },\; \Gamma : \mbox{weight 0} \ ,\quad
\partial_{\mu} : \mbox{weight 1} \ ,\quad
R,\; R_{\mu\nu} : \mbox{weight 2} \ ,\quad
{\delta \Gamma \over \delta \hat G_{\mu\nu}} : 
\mbox{weight $d$} \ .
\]
Using these rules and (\ref{HJ}), one obtains 
the equations, which depend on the weight as
\bea
\label{wt1}
\sqrt{\hat G}{\cal L}_{d} 
&=& \left[ \{ S_{loc},S_{loc} \} \right]_0 +
\left[ \{ S_{loc},S_{loc} \} \right]_2 \\
\label{wt2}
0 &=& \left[ \{ S_{loc},S_{loc} \} \right]_w  
\quad (w=4,6,\cdots d-2), \\
\label{wt3}
0 &=& 2\left[ \{ S_{loc}, \Gamma \} \right]_d 
+ \left[ \{ S_{loc},S_{loc} \} \right]_d
\eea
The above equations which determine 
$\left[ {\cal L}_{loc} \right]_{w}$. 
$\left[ {\cal L}_{loc} \right]_{0}$ and $[{\cal L}_{loc}]_{2}$ are
parametrized by 
\be
\label{vi}
\left[ {\cal L} _{loc} \right]_0 = W\ ,\quad 
\left[ {\cal L} _{loc} \right]_2 = -\Phi R \ .
\ee
Thus one can solve (\ref{wt1}) as
\bea
\label{vii}
\Lambda = {d \over 4(d-1)}W^{2} \ ,\quad 
1 = {d-2 \over 2(d-1)} W\Phi\ .
\eea
The case of $d=2$ is special and instead of 
Eqs.(\ref{wt1},\ref{wt2},\ref{wt3}), we obtain
\be
\label{d2ii}
\sqrt{\hat G}{\cal L}_2 = 
2\left[ \{ S_{loc}, \Gamma \} \right]_2 
+ \left[ \{ S_{loc},S_{loc} \} \right]_0 +
\left[ \{ S_{loc},S_{loc} \} \right]_2 \ .
\ee
When $d=2$, the second equation  
in (\ref{vii}) is irrelevant but by using (\ref{dSL}), 
we obtain
\be
\label{d2i}
W_2=- {2 \over l}\ . 
\ee
When $d>2$, by using (\ref{dSL}), one obtains $W$ and $\Phi$ as
\be
\label{viii}
W=-{2(d-1) \over l},\quad \Phi = -{l \over d-2}.
\ee
Note that there is an ambiguity in the choice of the 
sign but the relative sign of $W$ and $\Phi$ is different 
from AdS case. 
When $\left[ {\cal L}_{loc} \right]_{4}=0$, one gets
\bea
\label{ano4}
\lefteqn{{1 \over \sqrt{\hat G}}\left[\{ S_{loc},S_{loc} \} 
\right]_{4}} \nn
&&={dl^2 \over 4(d-1)(d-2)^{2}}
 R^{2}  - {l^2 \over (d-2)^{2}}R_{\mu\nu}R^{\mu\nu}\ . 
\eea
The weight $d$ flow equation (\ref{wt3}), which is related 
with the conformal anomaly in $d$ dimensions \cite{BVV,FMS}, is
written by
\bea
\label{ano1}
{W \over 2(d-1)}{1\over \sqrt{\hat G}}
\hat G_{\mu\nu}{\delta \Gamma \over \delta \hat G_{\mu\nu}}
= 2\left[ \{ S_{loc} , \Gamma \} \right]_{d} 
=-\left[ \{ S_{loc} , S_{loc} \} \right]_{d} .
\eea 
This $\hat G_{\mu\nu}{\delta \Gamma \over \delta \hat G_{\mu\nu}}$
can be regarded as the sum of conformal anomaly ${\cal W}_{d}$ 
and the total derivative term $\nabla_{\mu}{\cal J}^{\mu}_{d}$
in $d$ dimensions.  Thus we rewrite (\ref{ano1}) as following
\bea
\label{ix}
\kappa^2{\cal W}_{d}+\nabla_{\mu}{\cal J}^{\mu}_{d}
=-{d-1 \over W \sqrt{\hat G}}
\left[ \{S_{loc},S_{loc} \}\right]_{d} .
\eea
Here $\kappa^2$ is $d+1$ dimensional gravitational coupling.
Using the above relation, one can get the holographic conformal 
anomaly in 2 dimensions 
\be
\label{x}
\kappa^2{\cal W}_2 = -{l\over \sqrt{\hat G} }_2
\left[ \{ S_{loc} , \Gamma \} \right]_2
= {l \over 2}\hat R\ .
\ee
Actually, the corresponding result (via calculation
of central charge for corresponding 2d CFT) has been already obtained 
by Strominger\cite{strominger} using method similar to the one
developed in three-dimensional AdS space \cite{marc}.
It has been also proposed in ref.\cite{strominger} that holographic
conformal anomaly evaluation may be applied as well.
The result for 4d holographic conformal anomaly is obtained from
 (\ref{ano4}):
\be
\label{xi}
\kappa^2{\cal W}_{4} = {l\over 2\sqrt{\hat G} }
\left[ \{S_{loc},S_{loc} \}\right]_{4} 
= l^{3}\left( {1\over 24}\hat R^{2} -
{1\over 8} \hat R_{\mu\nu}\hat R^{\mu\nu}
\right)\ .
\ee
This agrees with the result for holographic conformal anomaly  
calculated  by various methods using AdS/CFT 
duality\cite{anom}. It indicates that all results about holographic
conformal anomaly \cite{anom} in AdS/CFT may be easily used in dS/CFT. 

We should note that holographic conformal anomaly obtained from 
dS/CFT duality seems to be identical with that from AdS/CFT one.
 This shows 
 that  obtained central charge from dS/CFT duality itself 
is the same with that from AdS/CFT. Nevertheless, it does not mean
that boundary CFTs should be necessarily the same because there may exist
 several 
different theories with the same central charge.  
Finally, let us note that the fact that holographic conformal anomaly 
from AdS/CFT or dS/CFT duality is the same suggests that both
these dualities are the consequence of some underlying fundamental
principle.
Even more, one can speculate on existance of more dualities of such sort,
also for other spaces.

\end{document}